\documentclass[%
 aip,
 jmp,%
 amsmath,amssymb,
 preprint,%
prl]{revtex4-1}

\usepackage{graphicx}% Include figure files
\usepackage{dcolumn}% Align table columns on decimal point
\usepackage{bm}% bold math
\usepackage{xcolor}

\begin{document}
\newcounter{subfigure}
%\preprint{AIP/123-QED}

\title[A Fractional Fokker-Planck Model for Anomalous Diffusion]{A Fractional Fokker-Planck Model for Anomalous Diffusion}% Force line breaks with \\
%\thanks{Footnote to title of article.}

\author{Johan Anderson}
\email{anderson.johan@gmail.com.}
\affiliation{ 
Department of Earth and Space Sciences, Chalmers University of Technology, SE-412 96 G\"{o}teborg, Sweden
}%
\author{Eun-jin Kim}%
\affiliation{ 
University of Sheffield, Department of Mathematics and Statistics, Hicks Building, Hounsfield Road, Sheffield, S3 7RH, UK
}%
\author{Sara Moradi}%
\affiliation{ 
Ecole Polytechnique, CNRS UMR7648, LPP, F-91128, Palaiseau, France
}%

\date{\today}% It is always \today, today,
             %  but any date may be explicitly specified

\begin{abstract}
In this paper we present a study of anomalous diffusion using a Fokker-Planck description with fractional velocity derivatives. The distribution functions are found using numerical means for varying degree of fractionality of the stable L\'{e}vy distribution. The statistical properties of the distribution functions are assessed by a generalized normalized expectation measure and entropy in terms of Tsallis statistical mechanics. We find that the ratio of the generalized entropy and expectation is increasing with decreasing fractionality towards the well known so-called sub-diffusive domain, indicating a self-organising behavior. 
\end{abstract}

\pacs{52.25.Dg, 52.30.Gz, 52.35.Kt}% PACS, the Physics and Astronomy
                             % Classification Scheme.
\keywords{Anomalous Diffusion, Fractional Fokker-Planck Equation, Self-Organisation}
\maketitle

\section{\label{sec:level1} Introduction}
In the early 20th century Einstein studied classical diffusion in terms of Brownian motion. In this process, the mean value of the process vanishes whereas the second moment or variance grows linearly with time $\langle \delta x^2\rangle = 2 D t$. Anomalous diffusion, however, is in contrast to classical diffusion in terms of the variance that exhibit a non-linear increase with time $\langle \delta x^2\rangle = 2 D t^{\alpha}$. There is no mechanism that inherently constrains $\lim_{\delta x, \delta t \rightarrow 0} \frac{\delta x^2}{\delta t}$ to be finite or non-zero. In more general terms, there are two limits of interest where the first is super-diffusion $\alpha > 1$ and the second is sub-diffusion with $\alpha < 1$. Such strange kinetics\cite{schlesinger1993, sokolov2002, klafter2005, metzler2000, metzler2004, mandelbrot1982} may be generated by accelerated or sticky motions along the trajectory of the random walk\cite{krommes2002}. The main cause of anomalous super-diffusion is the existence of long-range correlations in the dynamics generated by the presence of anomalously large particle displacements. 

Anomalous sub-diffusive properties has been studied in many different contexts, among them are that of holes in amorphous semiconductors where a waiting time distribution with long tails was introduced\cite{montroll1973} and the sub-diffusive processes within a single protein molecule described by generalized Langevin equation with fractional Gaussian noise\cite{kou2004}. Moreover, it has been recognized that the nature of the transport processes common to plasma physics is dominated by turbulence with a significant ballistic or non-local component where a diffusive description is improper. The basic mechanism underlying plasma transport is a very complex process and not very well understood. Furthermore, in plasma physics, super-diffusive properties are often found with $\alpha > 1$ such as the thermal and particle flux in magnetically confined plasmas or transport in Scrape-Off Layer (SOL) dominated by coherent structures\cite{carreras1996, carreras1999, milligen2005, negrete2005, sanchez2008}. In this paper, we will mainly concern ourselves with super-diffusion modeled by a Fractional Fokker-Planck equation (FFPE). 

A salient component describing the suggestive non-local features of plasma turbulence is the inclusion of a fractional velocity derivative in the Fokker-Planck (FP) equation leading to an inherently non-local description as well as giving rise to non-Gaussian probability density functions (PDFs) of, e.g., densities and heat flux. Note that, the non-Gaussian features of the PDFs heat or particle flux generated by non-linear dynamics in plasmas may be reproduced by a linear, though, fractional model. The non-locality is introduced through the integral description of the fractional derivative, and the non-Maxwellian distribution function drives the observed PDFs of densities and heat flux far from Gaussian \cite{anderson2010} as well as shear flow dynamics\cite{kim2009}. Some previous papers on plasma transport have used models including a fractional derivative where the fractional derivative is introduced on phenomenological premises\cite{sanchez2006}. In the present work we introduce the L\'{e}vy statistics into the Langevin equation thus yielding a fractional FP description. This approach is similar to that of Ref. \onlinecite{sanchez2006, moradipop2011, moradipop2012} resulting in a phenomenological description of the non-local effects in plasma turbulence. Using fractional generalizations of the Liouville equation, kinetic descriptions have been developed previously\cite{zaslavsky, zaslavsky2002, tarasov2005, tarasov2006}.

In investigations of the anomalous character of transport a useful tool is the non-extensive statistical mechanics which provides distribution functions intermediate to that of Gaussians and L\'{e}vy distributions adjustable by a continuous real parameter $q$\cite{tsallis1995lnp, tsallis1996, tsallis1998}. The parameter $q$ describes the degree of non-extensivity in the system. Non-extensive statistical mechanics has a solid theoretical basis for analysing complex systems out of equilibrium where the total entropy is not equal to the sum of the entropies from each subsystem. For systems comprised of independent or parts interacting through short-range forces the Boltzmann-Gibb statistical mechanics is sufficient however for systems exhibiting fractal structure or long range correlations this approach becomes unwarranted. Tsallis statistics is now widely applied e.g. to solar and space plasmas such as the heliosphere magnetic field and the solar wind\cite{balasis2011, pavlos2012, pavlos20122}.

Note that due to the obtained L\'{e}vy type distribution functions, higher moments will diverge thus it is of interest to define convergent statistical measures of the underlying process. We will employ the generalized $q$-moments or $q$-expectations as $\langle v^p \rangle_q = \int dv F(v)^q v^p/ \int dv F(v)^q$. The $q$-expectation can be a convergent moment of the distribution function although the regular moments diverges. This also gives us the opportunity to have a convergent pseudo-energy that is always convergent.  

The aim of this study is to elucidate on the non-extensive properties of the velocity space statistics and characterization of the fractal process in terms of Tsallis statistics. In Ref. \onlinecite{moradipop2011, moradipop2012} two limiting cases of the forced FFPE were studied using an expansion in the fractionality parameter $\alpha$ close to two. However, due to the intractability of these models a minimal model for the FFPE is thus constructed retaining mainly the effects of the fractional operator in order to understand the properties of the FFPE. Furthermore, in order to establish an effective connection to the extended statistical mechanics we obtain numerical solutions of the PDFs derived from the Tsallis statistic which is in good agreement with those found using the FFPE\cite{prato1999}. Furthermore, we consider numerical solutions to the Langevin system with L\'{e}vy distributed noise and show qualitatively similar results as the analytical work. Moreover, we find that self-organising behavior is present in the system where the ratio of the entropy and energy expectation is decreasing with decreasing fractionality. 

The remainder of the paper is organized as in Sec. II the Fractional Fokker-Planck Equation is introduced whereas in Sec. III a numerical study of the probability distribution functions obtained is presented. In Sec. IV, the resulting numerical entropies are discussed while the paper is concluded by results and discussions in Sec. V.

\section{\label{sec:level2} Fractional Fokker-Planck Equation}
The motion of a colloidal particle, i.e. Brownian motion, is described by a stochastic differential equation also known as the Langevin equation\cite{chandrasekhar}. It is assumed that the influence of the background medium can be split into a dynamical friction and a fluctuating part, $A(t)$, represented by Gaussian white noise. The Gaussian white noise assumption is usually imposed in order to obtain a Maxwellian velocity distribution describing the equilibrium of the Brownian particle. This connection is due to the relation between the Gaussian central limit theorem and classical Boltzmann-Gibbs statistics \cite{Khintchine}. However, the Gaussian central limit theorem is not unique and a generalization was done by L\'{e}vy \cite{levy}, Khintchine \cite{Khintchine} and Seshadri \cite{seshadri} by using long tailed distributions.

The underlying physical reasoning is to allow for the non-negligible probability of preferred direction and long jumps, i.e., L\'{e}vy flights, which therefore allows for asymmetries and long tails in the equilibrium PDFs, respectively. In the present work we introduce the L\'{e}vy statistics into the Langevin equation thus yielding a fractional FP description. Following the approach used by Barkai \cite{barkai} and Moradi \cite{moradipop2011, moradipop2012} a Fractional Fokker-Planck Equation (FFPE) with a fractional velocity derivative in the presence of a constant external force is obtained as
\begin{eqnarray}\label{eq:1.1}
\frac{\partial F}{\partial t}+\mathbf{v}\frac{\partial F}{\partial \mathbf{r}}+\frac{\mathbf{F}}{m}\frac{\partial F}{\partial \mathbf{v}}=\nu\frac{\partial }{\partial \mathbf{v}}(\mathbf{v}F)+D\frac{\partial^{\alpha} F}{\partial |\mathbf{v}|^{\alpha}},
\end{eqnarray}
where $0\le\alpha\le 2$. Here, the term $\frac{\partial^{\alpha} F}{\partial |\mathbf{v}|^{\alpha}}$ is the fractional Riesz derivative. The diffusion coefficient, $D$, is related to the damping term $\nu$, according to a generalized Einstein relation \cite{barkai} 
\begin{eqnarray}\label{eq:1.2}
D=\frac{2^{\alpha-1}T_{\alpha}\nu}{\Gamma(1+\alpha)m^{\alpha-1}}.
\end{eqnarray}
Here, $T_{\alpha}$ is a generalized temperature, and taking force $\mathbf{F}$ to represent the Lorentz force acting on the particles with mass $m$ and $\Gamma(1+\alpha)$ is the Euler gamma function. 

In order to analytically investigate the effects of the fractional derivative on the diffusion we consider the force-less homogeneous one dimensional Fokker-Planck equation of the form,
\begin{eqnarray}\label{eq:1.3}
\frac{\partial F}{\partial t} = \nu \frac{\partial }{\partial v} (vF) + D \frac{\partial^{\alpha}}{\partial |v|^{\alpha}} F.
\end{eqnarray}
The solution is found by Fourier transforming and treating the fractional derivative in the same manner as in Ref. \citep{saichev1997} we find,
\begin{eqnarray}\label{eq:1.4}
\frac{\partial \hat{F}}{\partial t} = - \nu k \frac{\partial }{\partial k} \hat{F} - D |k|^{\alpha} \hat{F}.
\end{eqnarray}
The stationary PDF is now readily obtained by integration and an inverse Fourier transform,
\begin{eqnarray}
\hat{F}(k) & = & F_0 \exp{(- \frac{D}{\nu \alpha} |k|^{\alpha})} \label{eq:1.5}, \\
F(v) & = & \frac{F_0}{2 \pi}\int_{-\infty}^{\infty} dk \exp{ (- \frac{D}{\nu \alpha} |k|^{\alpha} + i kv)}. \label{eq:1.6}
\end{eqnarray}

Due to the fractal form of the inverse Fourier transform analytical solutions of the PDF for the general case is difficult to obtain, except in particular cases of $\alpha = 1.0$ and $\alpha = 2$ yielding a Lorentz distribution and a Gaussian distribution, respectively. Note, that Eq. (\ref{eq:1.6}) is equivalent to what was found in Ref. \onlinecite{tarasov2006}. From a different perspective, for a PDF of a single variable the Tsallis statistics may be generated by an appropriately constructed Langevin equation of the form,

\begin{equation}
\frac{d v}{dt} = K(v) + \frac{dD(v)}{dv} + \sqrt{2 D(v)} w(t). \label{eq:2.1}
\end{equation}
This result was obtained in Ref. \onlinecite{lutz2003}, where $K(v)$, $D(v)$ and $dD(v)/dv$ are non-stochastic functions of $v$ and $w(t)$ is white-in-time Gaussian noise. The PDF generated by Eq. (\ref{eq:2.1}) is,
\begin{equation}
F(v) = \frac{N}{(1 + \beta (q-1)v^2)^{1/(q-1)}}. \label{eq:2.2}
\end{equation}
Note that $q>1$ and $\beta$ are found from the analytical forms of $D(v)$ and $K(v)$ as well as the $v^2$ dependence. Here $N$ is normalization factor. Furthermore, it is found that $\beta$ is not representative of an inverse temperature of the system due to its non-equilibrium nature. It has been recognized that multi-fractal models stemming from the Tsallis statistical mechanics may well describe isotropic fluid turbulence at high but finite Reynolds number \cite{gotoh2004}. In the next Section we will study the solutions to the Eq. (\ref{eq:1.6}) in more detail.
\section{\label{sec:level3} Numerical solutions to the fractional Fokker-Planck Equation}
The main topic of this paper is to evaluate the statistical properties in terms of Tsallis statistics dependent of the fractional index ($\alpha$) in Eq. (\ref{eq:1.6}). We will start by numerically computing the PDFs with $\alpha$ as our free parameter, and next we will fit the computed PDFs to the proposed generalized analytical Cauchy-Lorentz PDFs found from Tsallis statistical mechanics. Subsequently, in order to statistically evaluate the numerically found PDFs in the fractal model we will determine the $q$-expectation and the Tsallis non-extensive entropy. Note that the regular statistical moments of the PDFs will not converge unless the PDFs are considered to have a finite compact support. Although, the PDFs have been studied before in Ref.\cite{tsallis1995, prato1999} we show the PDFs and fits to the analytically found PDFs for completeness and for reference during the $q$-entropy study. Thus, we will now focus on solving Eq. (\ref{eq:1.6}) numerically, by computing the inverse Fourier transform and compare the found PDFs to previously derived analytical solutions.

In Figure 1, the numerically found PDFs are shown (log-linearly) for $\alpha = 0.25$ (black dashed line), $\alpha = 0.50$ (cyan line), $\alpha = 0.75$ (yellow line), $\alpha = 1.00$ (magenta line), $\alpha = 1.25$ (green line), $\alpha = 1.50$ (red line), $\alpha = 1.75$ (blue line) and $\alpha = 2.00$ (black line). Here, in this study the diffusion coefficient over the dissipation is kept constant $D/\nu = 1.0$. 

We note as the parameter $\alpha$ decreases, the normalized fourth moment (Kurtosis = $m_4/m^2_2$ = the ratio of the fourth moment divided by the square of the standard deviation) of the symmetric PDF increases rapidly where PDFs become more and more peaked with elevated tails. Note that the distribution varies smoothly as $\alpha$ - parameter is decreased from a Gaussian distribution with $\alpha = 2.0$ passing through the Lorentz distribution with $\alpha=1.0$.

In line with the stochastic non-linear analysis presented in Eqs (\ref{eq:2.1}) - (\ref{eq:2.3}) it has been shown, in Ref. \onlinecite{tsallis1995}, that using generalized statistical mechanics yielded PDFs that are of Cauchy-Lorentz form,
\begin{eqnarray}
F(v) = \frac{a}{(1 + b (q-1) v^2)^{1/(q-1)}}.  \label{eq:1.7}
\end{eqnarray}
We note that this type of PDF exhibit power law tails that are significantly elevated compared to Gaussian or exponential tails, c.f. Eq. (\ref{eq:2.1}). It is worthy noting that the precise analytical relation between the fractality index $\alpha$ and the non-extensivity parameter $q$ is not entirely clear. One possibility is the formal relation between the fractality index $\alpha$ and the non-extensivity $q$ proposed by  \cite{tsallis1995} as:
\begin{equation}
\alpha = \frac{3-q}{q-1}. \label{eq:2.4}
\end{equation}
A second possibility is the following simple relation in \cite{tsallis1995}.
\begin{equation}
\alpha = \frac{1}{q-1}. \label{eq:2.3}
\end{equation}
Taking Gaussian limit in Eq. (\ref{eq:2.3}) requires a caution as it cannot reproduce the limit of  $\alpha = 2.0$ where $q=1$. Interestingly, comparing Eq. (\ref{eq:2.3}) there is a direct connection between non-linear dynamics and the fractional FP model. However, Eq. (\ref{eq:2.3}) yields a seemingly erroneous scaling of the tail of the PDF. It can easily be shown that $F(v) \propto v^{-(\alpha + 1)}$ as $v \rightarrow \infty$ which is only fulfilled by Eq. \ref{eq:2.4} whereas \ref{eq:2.3} yields a steeper slope for the tails with $\alpha$. On the other hand in the limit of small $v$ the exponential factor can be approximated as follows $e^{ikv} = 1 +  ikv - \frac{1}{2} k^2 v^2 + \ldots$, keeping only the even powers in the integral due to the symmetry, we find:
\begin{eqnarray}
F(v) \approx \frac{F_0}{\pi} \frac{1}{\alpha} \left[ \left(\frac{D}{\alpha \nu}\right)^{-1/{\alpha}} \Gamma(\frac{1}{\alpha}) - \left( \frac{D}{\alpha \nu}\right)^{-3/{\alpha}} \Gamma(\frac{3}{\alpha}) \frac{v^2}{2!} + \ldots \right]. \label{eq:2.5}
\end{eqnarray}
Here $\Gamma$ is the gamma function. Note that for large $\alpha$ the relations from Eq.s (\ref{eq:2.4}) and (\ref{eq:2.3}) yield similar values and asymptotically approaches 1. Moreover, the last expansion in Eq. \ref{eq:2.5} approximates a Gaussian distribution function for small $v$ for both Eq.s (\ref{eq:2.4}) and (\ref{eq:2.3}).

To further investigate the suitability of Eqs. (10)-(11), in Figure 2, the PDFs are fitted using Eqs. (\ref{eq:2.3}) with $q = 5/3$ and utilizing (\ref{eq:2.4}) with $q=9/5$ for $\alpha = 3/2$. We find good agreement over several orders of magnitude between the proposed analytically derived PDFs based on Eqs. (\ref{eq:2.3}) and (\ref{eq:2.4}) and the numerically computed PDFs for the values used. Note that similar agreement is found for all values of $0.25 < \alpha < 1.5$. We find that using $q=5/3$ gives a slightly better fit for small $v$ compared to the higher $q = 9/5$ value whereas the tails seems to be off using Eq. (\ref{eq:2.3}). Since the tail parts are of great importance, we will thus use the relation in Eq. (\ref{eq:2.4}) throughout the rest of the paper.

While we follow the definition that any diffusive process that diverges from the form $\langle x^2 \rangle (t) \propto  t $ is called anomalous, in most cases we will deal with super-diffusion where $\langle x^2 \rangle (t)$ may be divergent. In order to find a useful statistical measure of the super-diffusive or fractal process we introduce,
\begin{eqnarray}
\langle v^2 \rangle_q = \frac{\int_{-\infty}^{\infty} dv (F(v))^q v^2}{\int_{-\infty}^{\infty} dv (F(v))^q},   \label{eq:1.8}
\end{eqnarray}
which we will call the $q$-expectation according to Ref. \onlinecite{prato1999}. Note, that e.g. the exactly solvable case with $\alpha = 1.0$ we find that the ordinary expectation diverges, however as $q$ increases a finite measure is found. Moreover, this gives also the opportunity to define a pseudo-energy in the system as the smallest possible value $q$ where the $q$-expectation converges. Naturally this reduces to the classical energy for $\alpha=2$.

In principle, all values of $v$ $F(|v| < \infty)$ should be used for the $q$-expectation of F(v). However for numerical tractability we have used a PDF with finite support $F(v)_{num} = F(v)$ for $|v| < 10 $ and zero everywhere else. Different support ranges have been tested where extending the range $|v| < 15 $ makes only minor changes.

Figure 3 shows the $q$-expectation as a function of $q$, treating $q$ as a free parameter for $\alpha = 0.5$ (magenta line), $\alpha = 1.0$ (blue line), $\alpha = 1.5$ (red line) and $\alpha = 2.0$ (black line). We find that just as expected the $q$-expectation falls off with $q$, however results corresponding to smaller $\alpha$ (more intermittent) falls off faster than the $q$-expectation of the Gaussian process. In general, there exist a smallest value for $q$ where the $q$-expectation is finite. In the case of $\alpha = 2.0$ it converges for all $q$ and we find that the PDF is a Gaussian with a variance $\sigma^2 = 1.0$.

In Figure 4, we have used the relation in Eq. (\ref{eq:2.4}) to determine the appropriate values for the different $\alpha$-stable processes of the PDFs determined in Figure 1, showing a scan from $q \in (1.5, 2.6)$ with a vertical line at $q=5/3$. We find that the $q$-expectation has a minimum at $q=5/3$ which is equivalent to value of $\alpha = 2.0$. This further confirms that that Eq. (\ref{eq:2.4}) is a suitable choice and also that  the $q$-expectation is a useful measure.

Furthermore, in Figure 5, the 1-expectation or energy is shown as a function (numerical result in black dots and a fit in red diamonds) of $\alpha$ utilizing the finite support of the PDFs shown in Figure 1. It is found that the 1-expectation decreases exponentially with increasing $\alpha$. Here we also note that the energy is convergent for all $\alpha$ due to the finite support of the PDFs. The main motivation for defining the $q$-expectation or tempered pseudo-energy is the sharp increase in the second moment for small $\alpha$, where the expectation value is diverging due to the long tails of the PDFs.

\section{\label{sec:entropy} Entropy}
In thermodynamics a measure on the number of ways a system can be arranged is denoted entropy. In terms of generalized statistical mechanics $q$-entropy or Tsallis entropy can be introduced as,
\begin{eqnarray}
S_q = \frac{1-\int dv (F(v))^q}{q-1}.   \label{eq:1.11}
\end{eqnarray}
Note that for Gaussian statistics the $q$-entropy is reduced (by L'Hospital's rule) to the conventional Boltzmann-Gibbs entropy $S = - \int dv \ \log(F(v)) F(v)$. Moreover, an important distinct property of the Tsallis generalized entropy is its non-extensivity, i.e. for two systems $A$ and $B$, the total entropy is not the sum of the entropies of the individual systems, $S_q(A+B) \neq S_q(A) + S_q(B)$. We will now use the Tsallis entropy to investigate the importance of fractal structure in velocity space and we will contrast the resulting generalized entropies to the standard Boltzmann-Gibbs entropy. 

In Figure 6, the dependence of the $q$-entropy as a function of $q$ for $\alpha = 0.5$ (magenta line), $\alpha = 1.0$ (blue line), $\alpha = 1.5$ (red line) and $\alpha = 2.0$ (black line) with $q$ as a free parameter is displayed. The $q$-entropy is rapidly decreasing with increasing $q$,  mainly due to the fact that the entropy is dependent on the $q$-th power of the PDF. The scaling of the $q$-entropy of $q$ using relation (\ref{eq:2.4}) is displayed in Figure 7 with a vertical line at $q=5/3$. Using the appropriate relation between $\alpha$ and $q$ shows that a maximum in the entropy is found for Gaussian statistics. For comparison, the Boltzmann-Gibbs entropy is shown in Figure 8 as a function of $q$ using the relation between $\alpha$ and $q$ as in Eq. \ref{eq:2.4}. We find that the entropy is monotonously increasing with increasing $q$. This is an indication that the Tsallis entropy is viable measure identifying the process with highest entropy as the Gaussian process. As an interesting measure of the dynamics and the importance of fractionality in the dynamics is to normalize the entropy with the energy, we will now discuss this quantity.

The $q$-entropy normalized with the $q$-expectation as a function of $q$ is displayed in Figure 9 for $\alpha = 1/2$ (magenta rings), $\alpha = 1.0$ (blue diamonds), $\alpha = 3/2$ (red triangles) and $\alpha = 2.0$ (black stars). The normalized $q$-entropy is rapidly increasing with increasing $q$ mainly due to the rapid decrease of the $q$-expectation ($\langle v^2\rangle_q$) with increasing $q$. This indicates that in a statistical mechanics sense the normalized generalized entropy is increasing with increasing $q$ as a process in velocity space in the sub-diffusive domain whereas in the range of small $\alpha$ the high-velocity, small likelihood events are more dominant. In the sub-diffusive regime the small amplitude events are dominant. In Figure 10, the $q$-entropy normalized to the $q$ as a function of $q$ determined by the relation \ref{eq:2.4} is displayed. We find that the maximum is found at the same $q=5/3$ values as the vertical line. Furthermore in contrast in Figure 11 the Boltzmann-Gibbs entropy normalized to the energy (black dots) and the $q$-expectation (blue stars) as a function of $q$ with a vertical line at $q=5/3$ is shown. Here the $q$ values are determined by Eq. (\ref{eq:2.4}). The normalized Boltzmann-Gibb entropy increases as a function of $q$ when normalized to the regular energy due to the influence of the strong tails. Also in the normalized description the $q$-entropy can identify the Gaussian process as the process with highest normalized entropy. However, the entropy normalized to the $q$-expectation is first increasing for small $q$ in the sub-diffusive region and then decreases for larger $q$ in the super-diffusive region. 
The dynamics of the FFPE system is dependent on the level of diffusion in comparison to the collisional damping in terms of the ratio of the two free parameters $D/\nu$ c.f. Eq. \ref{eq:1.6}. Allowing for a controlled variation of the free parameter $D/\nu$ is an important aspect on elucidating on the complex fractional dynamics. Note that, the width of the PDFs are significantly dependent on this factor, making the normalization with $\langle v^2\rangle$ and $\langle v^2\rangle_q$ impossible. In Figure 12 and 13 the q-entropy and Boltzmann-Gibbs entropy are shown with $D/\nu$ as a parameter, respectively. In this study the discrete values of $D/\nu$ is chosen to be in the range $\{ 10^{-2}, 10^{-1}, 1, 10^{1}, 10^{2}\}$. In Figure 12, it is found that the q-entropy is decreasing with $q$ for small $D/\nu$ (collision dominated) whereas the opposite or a flat profile is found for large $D/\nu$ (diffusion dominated). Furthermore, in Figure 13 the Boltzmann-Gibbs entropy is presented where the entropy is increasing with increasing $q$, in particular in the diffusion dominated regime. 
In the above it is indicated that as the fractality index increases (smaller $q$) the entropy decreases indicating a self-organising behavior in velocity space with long-range correlations. The number of possible microscopical realizations decreases fast. 

As a comparison to the analytical work, we numerically study the influence of L\'{e}vy stable processes as a noise source in the Langevin equation of motion. 
\begin{eqnarray}
\frac{d x}{dt} & = & v \\
\frac{d v}{dt} & = & -\nu v + \zeta.
\end{eqnarray}
Here the variables are normalized in the numerical scheme as $\tau = \nu t$ and $\zeta$ is the stochastic forcing. In the simulations we follow $100000$ particles until the system comes to a quasi-steady state. The numerical PDFs are determined by using 200 bins and we consider variations of the fractionality between $\alpha \in \{1.1,..,2.0\}$. Note that also here the PDFs are found to fall off as $F(v) \sim v^{-\alpha - 1}$. The resulting Boltzmann-Gibbs and Tsallis entropies are given in Figure 14. Contrasting the results with those shown in Figure 7 and 8, it is found that the Boltzmann-Gibb entropy is increasing with $q$ and that the Tsallis entropy is almost flat in the current range and finally seems to decrease for larger $q$. This is in good qualitative agreement with the analytical model however there are some quantitative differences. Note that the range $q$ is smaller compared to the analytical work.

\section{\label{sec:level4} Results and discussion}
Non-linear processes with non-Gaussian character have attracted significant attention during recent years calling for an efficient model describing such dynamics. In this paper we have investigated one prominent candidate capturing the main features in the dynamics, namely the Fractional extended Fokker-Planck Equation (FFPE). The FFPE is obtained by modifying the velocity derivative to a fractional differential operator allowing for non-local effects in velocity space. The underlying physical reasoning for using the FFPE is to allow for the non-negligible probability of direction preference and long jumps, i.e., L\'{e}vy flights, which therefore allows for asymmetries and long tails in the equilibrium PDFs, respectively.  

The aim of this study was to shed light on the non-extensive properties of the velocity space statistics and characterization of the fractal processes of the FFPE in terms of Tsallis statistics. The non-extensive statistical mechanics of Tsallis provides velocity space distribution functions intermediate to that of Gaussians and L\'{e}vy distributions adjustable by a continuous real parameter $q$ which seems to be suitable for comparing with the distribution found in FFPE. The parameter $q$ describes the degree of non-extensivity in the system. Non-extensive statistical mechanics has a solid theoretical basis for analysing complex systems out of equilibrium. For systems comprised of independent or parts interacting through short-range forces the Boltzmann-Gibb statistical mechanics is sufficient however for systems exhibiting fractal structure or long range correlations this approach becomes unwarranted.

In this work we have utilized generalized $q$-moments or $q$-expectations as $\langle v^p \rangle_q = \int dv F(v)^q v^p / \int dv F(v)^q$ as due to the obtained L\'{e}vy type character of the distributions, higher moments may diverge. The $q$-expectation results in  convergent moment although the regular moments diverges. This also permits us to define a convergent pseudo-energy that is always convergent. We show that the PDFs derived from the Tsallis statistic is in good agreement with those found using the FFPE. Moreover, we find that self-organising behavior is present in the system where the ratio of the entropy and energy expectation is decreasing with decreasing fractionality or increasing $\alpha$. 
 
Finally, it seems that a FFPE is a viable candidate for explaining certain non-linear features ubiquitous to anomalous plasma transport as well as for other physical processes. Note that in Ref. \onlinecite{gotoh2004}, a relation between Tsallis statistical mechanics and Navier-Stokes turbulence was established. A direct numerical comparison between the Langevin approach and the FFPE using Tsallis statistical mechanics is a possible topic for future work.

\section{\label{sec:level5} Acknowledgements}
The authors are grateful to the participants of Festival de th\'{e}orie 2013, organized by the C.E.A of France, for many valuable discussions.

%\section*{References}

\section{Figures}
\newpage
%%%%%%%%%%%%%bild%%%%%%%%%%%%%%%%%%%%
\begin{figure}[tbp]
\begin{center}
\includegraphics[width=10cm, height=8cm]{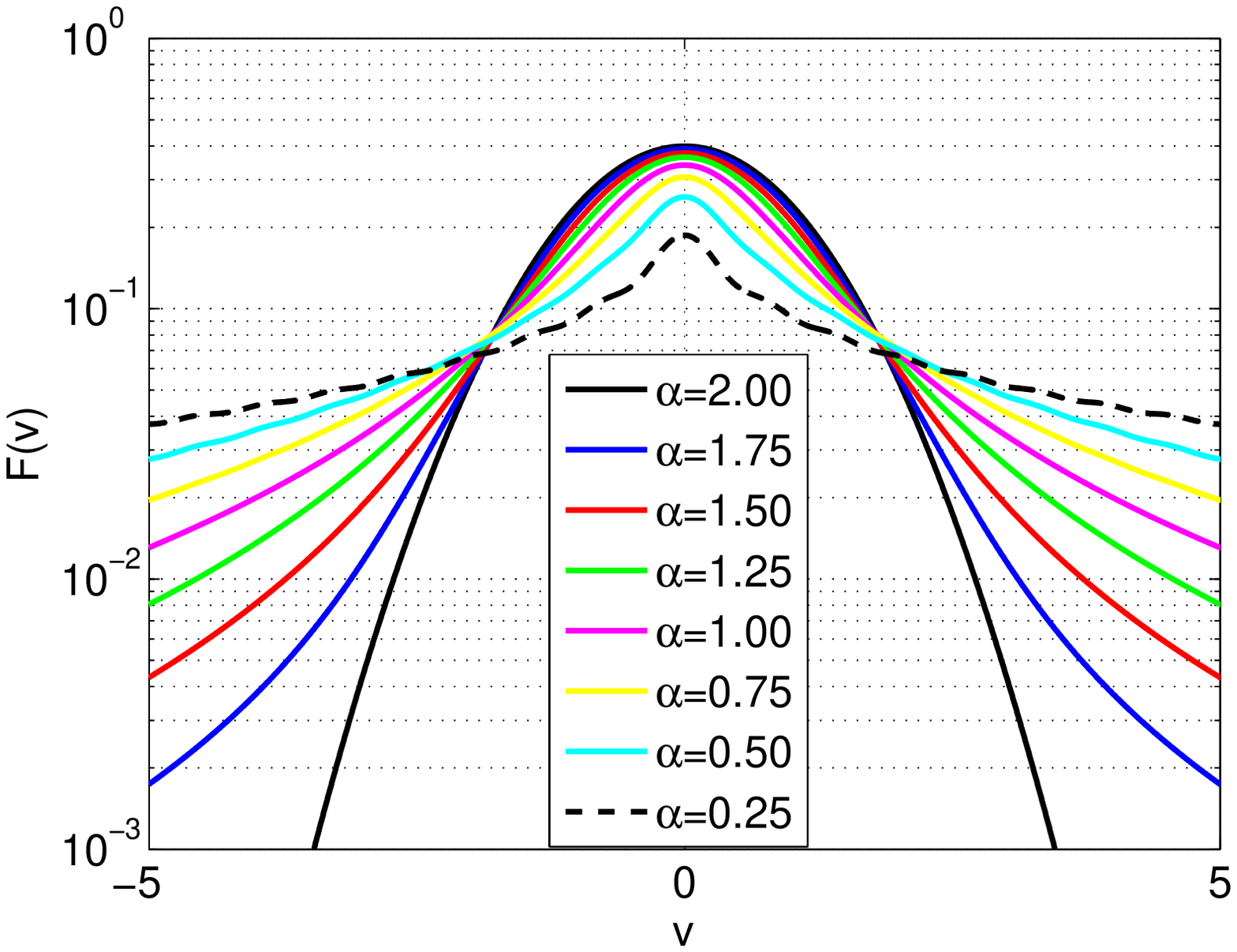}
\end{center}
\caption{The F(v) as a function of the velocity v for $\alpha = 2.00$ (black line), $\alpha = 1.75$ (blue line), $\alpha = 1.50$ (red line), $\alpha = 1.25$ (green line), $\alpha = 1.00$ (magenta line), $\alpha = 0.75$ (yellow line), $\alpha = 0.50$ (cyan line) and $\alpha = 0.25$ (black dashed line).}
\label{fig1}
\end{figure}
%%%%%%%%%%%%%bild%%%%%%%%%%%%%%%%%%%%

%%%%%%%%%%%%%bild%%%%%%%%%%%%%%%%%%%%
\begin{figure}[tbp]
\begin{center}
\includegraphics[width=10cm, height=8cm]{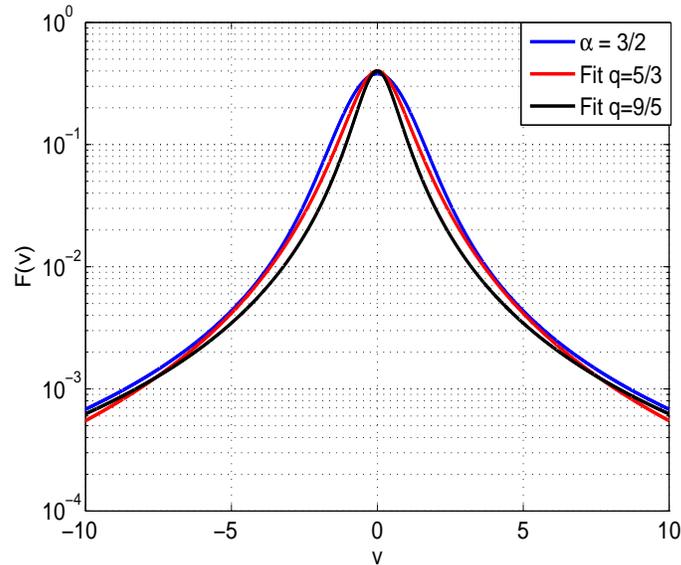}
\end{center}
\caption{The PDF as a function of the velocity $v$ for $\alpha = 1.5$ (blue line) fitted with  $q = 5/3$ (red line) and $q = 9/5$ (black line).}
\label{fig2}
\end{figure}
%%%%%%%%%%%%%bild%%%%%%%%%%%%%%%%%%%%

%%%%%%%%%%%%%bild%%%%%%%%%%%%%%%%%%%%
\begin{figure}[tbp]
\begin{center}
\includegraphics[width=10cm, height=8cm]{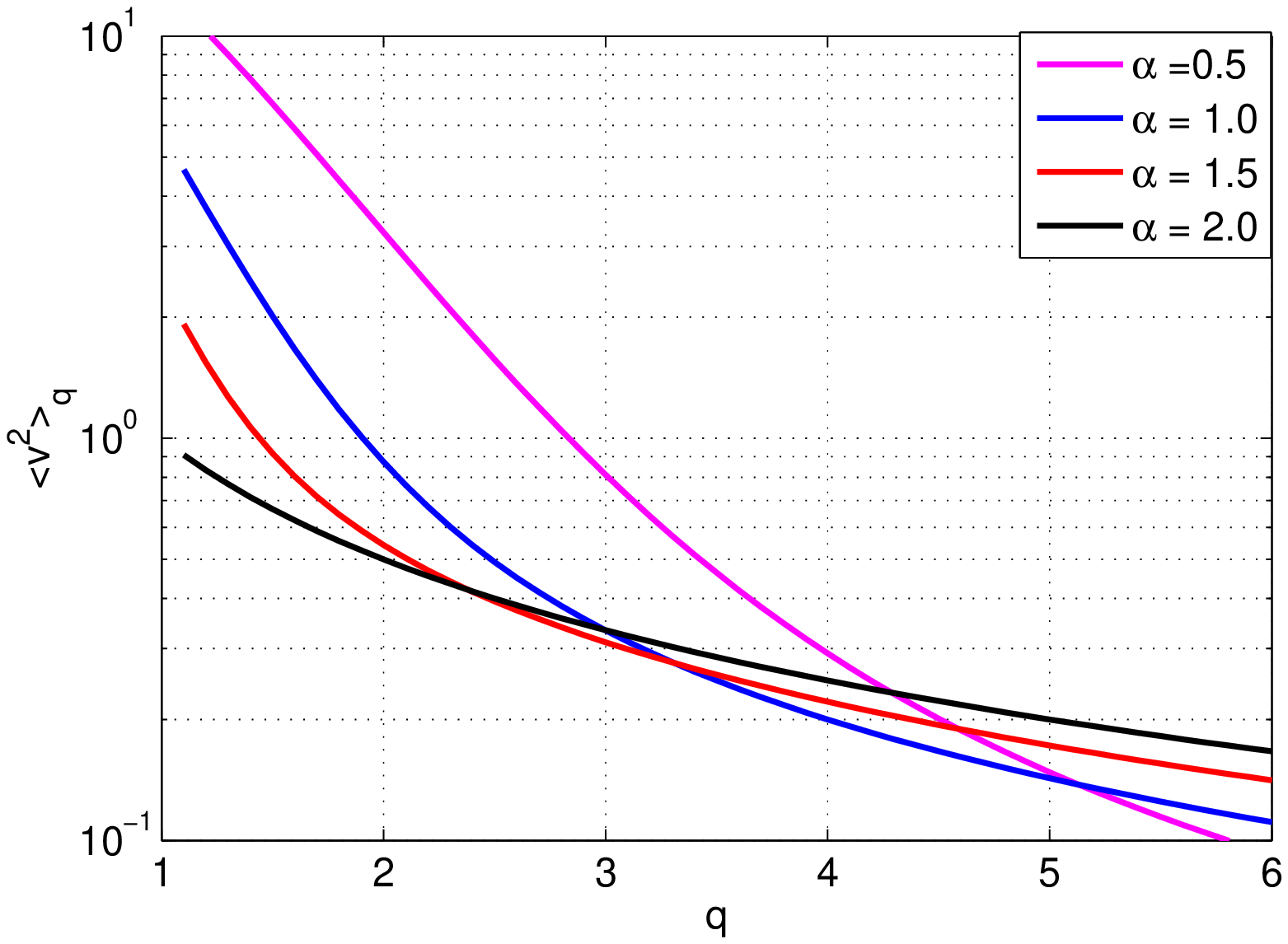}
\end{center}
\caption{The normalized $q$-expectation as a function of $q$ for $\alpha = 0.5$ (magenta line), $\alpha = 1.0$ (blue line), $\alpha = 1.5$ (red line) and $\alpha = 2.0$ (black line).}
\label{fig4a}
\end{figure}
%%%%%%%%%%%%%bild%%%%%%%%%%%%%%%%%%%%

%%%%%%%%%%%%%bild%%%%%%%%%%%%%%%%%%%%
\begin{figure}[tbp]
\begin{center}
\includegraphics[width=10cm, height=8cm]{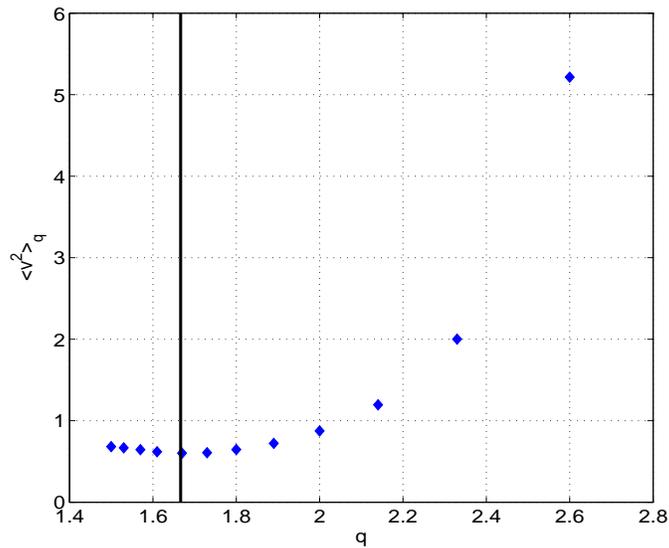}
\end{center}
\caption{The normalized $q$-expectation (blue stars) as a function of $q$ using Eq. \ref{eq:2.4} as a relation between the non-extensivity and the fractality with a vertical line at $q = 5/3$.}
\label{fig4b}
\end{figure}
%%%%%%%%%%%%%bild%%%%%%%%%%%%%%%%%%%%

%%%%%%%%%%%%%bild%%%%%%%%%%%%%%%%%%%%
\begin{figure}[tbp]
\begin{center}
\includegraphics[width=10cm, height=8cm]{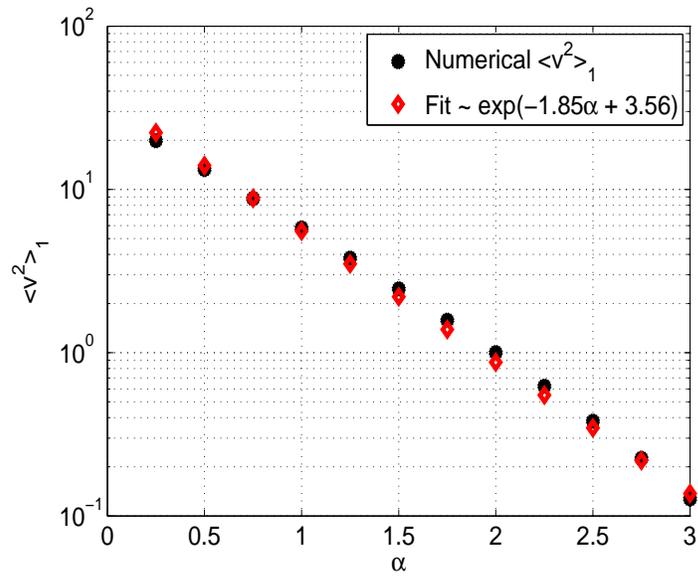}
\end{center}
\caption{The 1-expectation as a function of $\alpha$ for the numerically bounded PDFs with the numerical result in black dots and a fit in red diamonds.}
\label{fig5}
\end{figure}
%%%%%%%%%%%%%bild%%%%%%%%%%%%%%%%%%%%

%%%%%%%%%%%%%bild%%%%%%%%%%%%%%%%%%%%
\begin{figure}[tbp]
\begin{center}
\includegraphics[width=10cm, height=8cm]{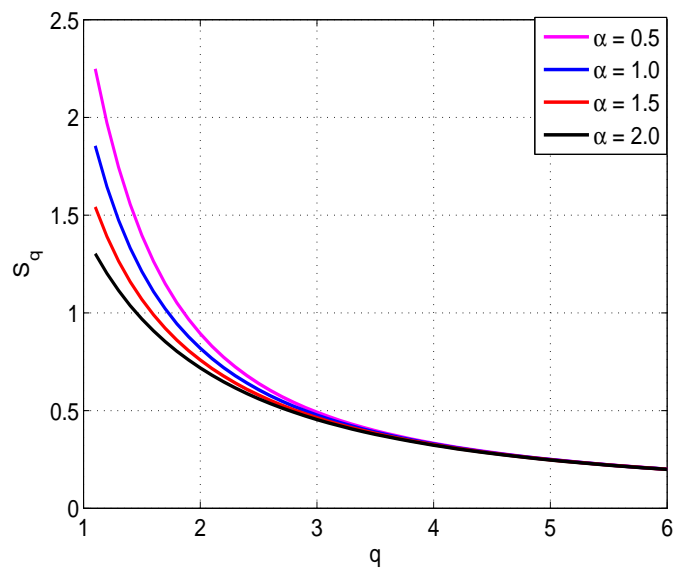}
\end{center}
\caption{The $q$-entropy as a function of $q$ for PDFs with $\alpha = 0.5$ (magenta line), $\alpha = 1.0$ (blue line), $\alpha = 1.5$ (red line) and $\alpha = 2.0$ (black line).}
\label{fig6a}
\end{figure}
%%%%%%%%%%%%%bild%%%%%%%%%%%%%%%%%%%%

%%%%%%%%%%%%%bild%%%%%%%%%%%%%%%%%%%%
\begin{figure}[tbp]
\begin{center}
\includegraphics[width=10cm, height=8cm]{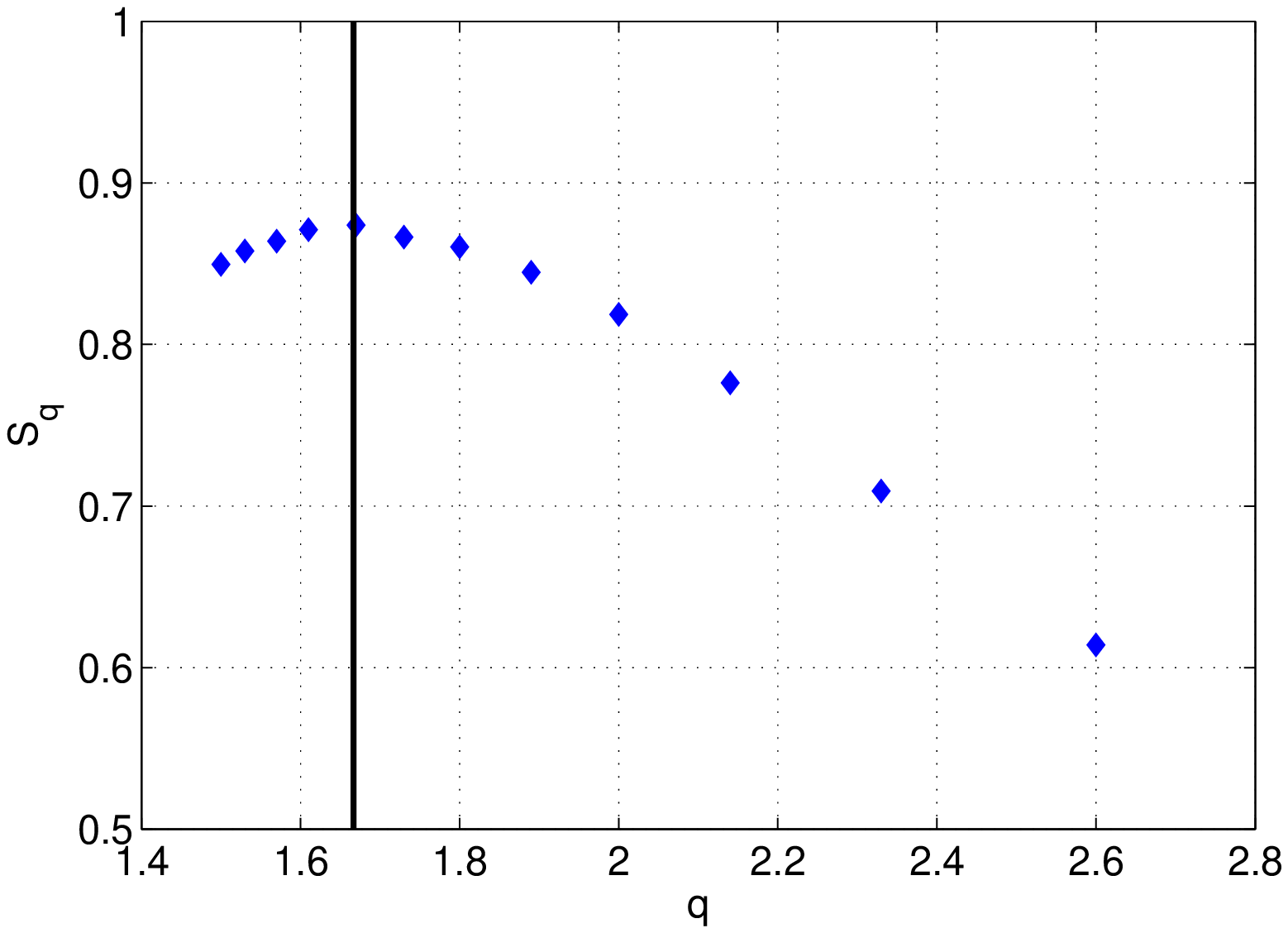}
\end{center}
\caption{The $q$-entropy as a function of $q$ using Eq. \ref{eq:2.4} as a relation between the non-extensivity and the fractality.}
\label{fig6b}
\end{figure}
%%%%%%%%%%%%%bild%%%%%%%%%%%%%%%%%%%%

%%%%%%%%%%%%%bild%%%%%%%%%%%%%%%%%%%%
\begin{figure}[tbp]
\begin{center}
\includegraphics[width=10cm, height=8cm]{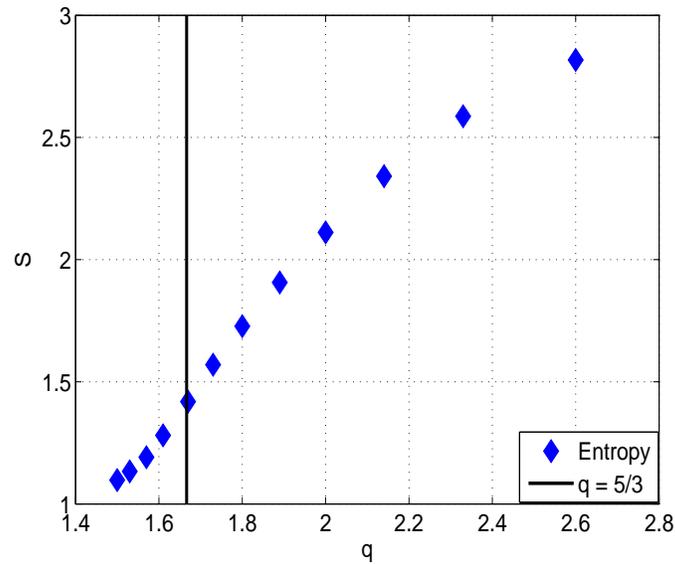}
\end{center}
\caption{The Boltzmann-Gibb entropy as a function of $q$ with a vertical line $q=5/3$.}
\label{fig6c}
\end{figure}
%%%%%%%%%%%%%bild%%%%%%%%%%%%%%%%%%%%

%%%%%%%%%%%%%bild%%%%%%%%%%%%%%%%%%%%
\begin{figure}[tbp]
\begin{center}
\includegraphics[width=10cm, height=8cm]{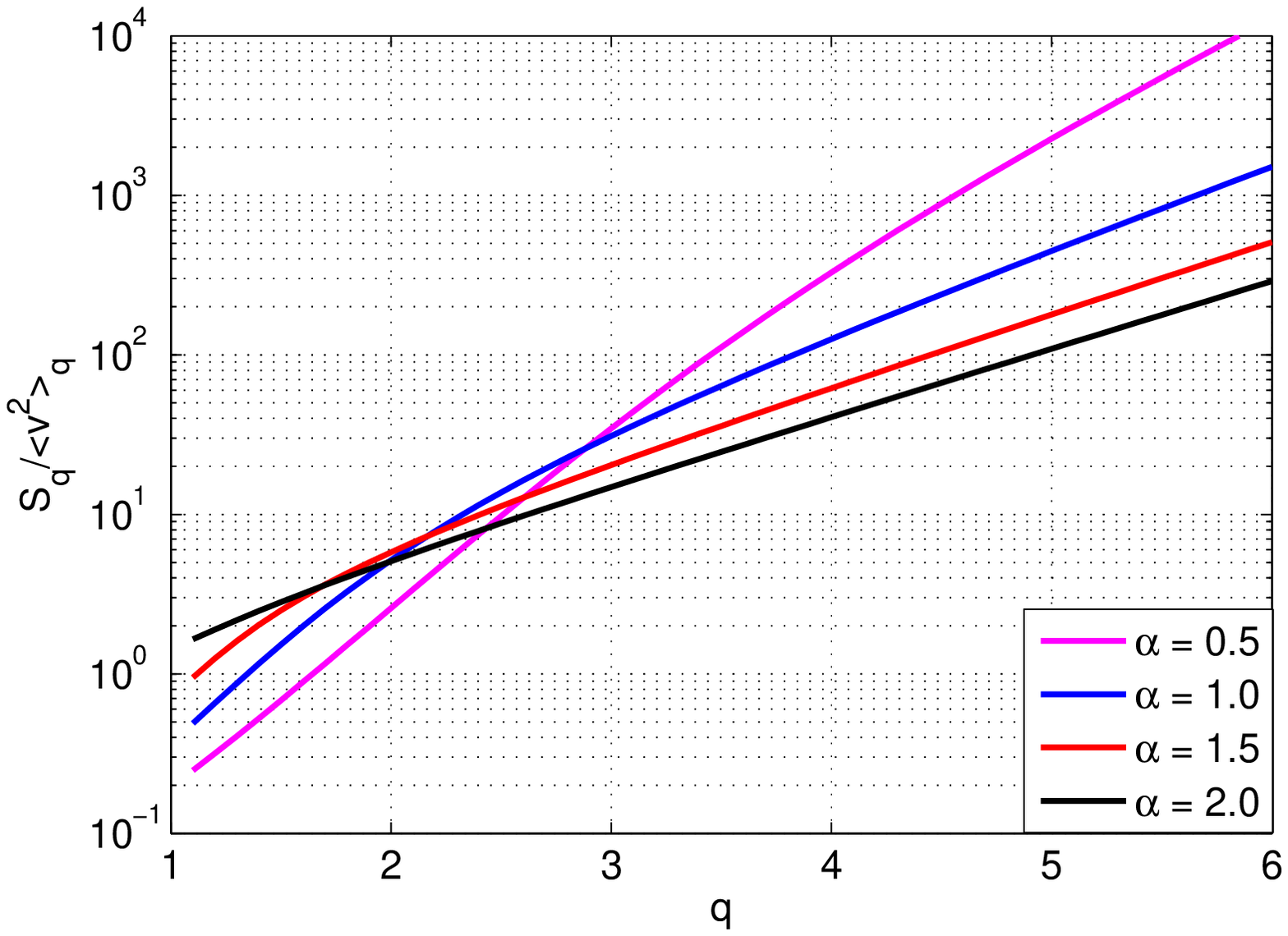}
\end{center}
\caption{The $q$-entropy normalized to the $q$-expectation $\langle v^2\rangle_q$ as a function of $q$ for PDFs with $\alpha = 0.5$ (magenta line), $\alpha = 1.0$ (blue line), $\alpha = 1.5$ (red line) and $\alpha = 2.0$ (black line).}
\label{fig7a}
\end{figure}
%%%%%%%%%%%%%bild%%%%%%%%%%%%%%%%%%%%

%%%%%%%%%%%%%bild%%%%%%%%%%%%%%%%%%%%
\begin{figure}[tbp]
\begin{center}
\includegraphics[width=10cm, height=8cm]{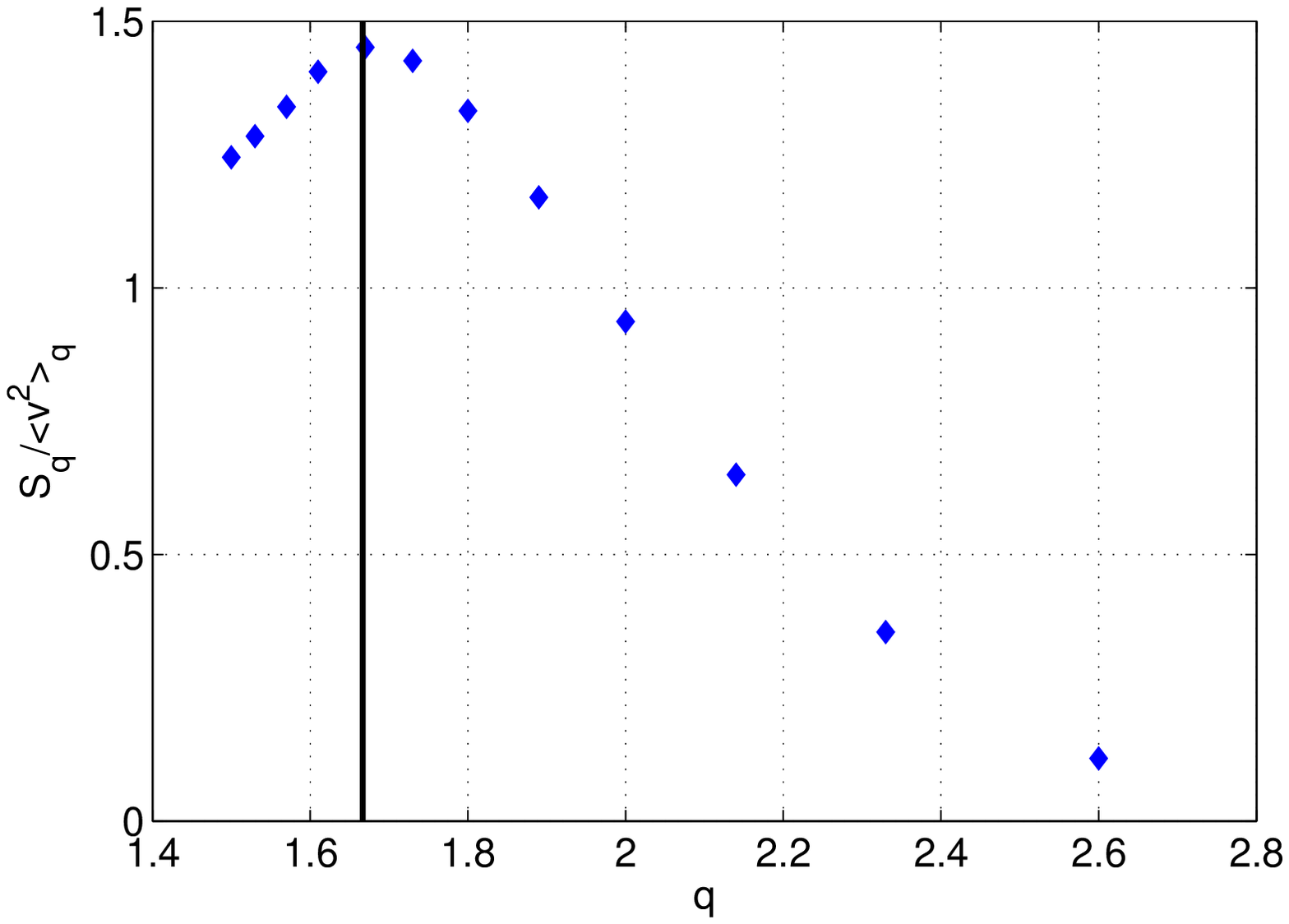}
\end{center}
\caption{The $q$-entropy normalized to the $q$-expectation $\langle v^2\rangle_q$ as a function of $q$ using Eq. \ref{eq:2.4} as a relation between the non-extensivity and the fractality with a vertical line at $q=5/3$.}
\label{fig7b}
\end{figure}
%%%%%%%%%%%%%bild%%%%%%%%%%%%%%%%%%%%

%%%%%%%%%%%%%bild%%%%%%%%%%%%%%%%%%%%
\begin{figure}[tbp]
\begin{center}
\includegraphics[width=10cm, height=8cm]{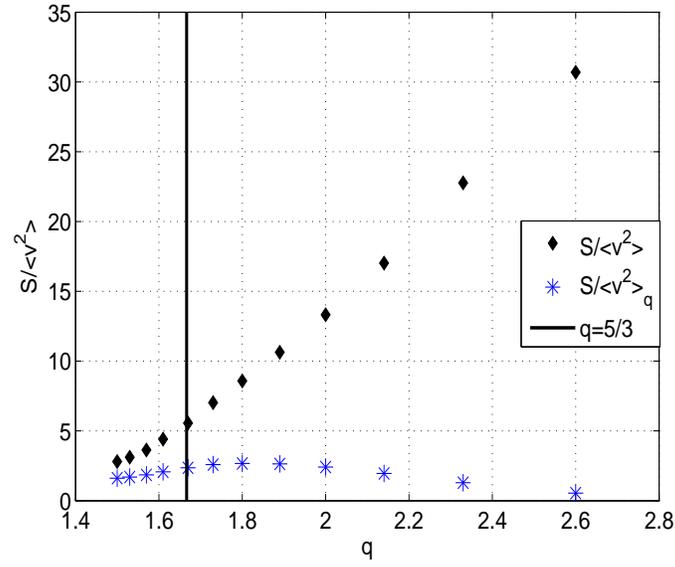}
\end{center}
\caption{The Boltzmann-Gibbs entropy normalized to the $q$-expectation $\langle v^2\rangle_q$ and the normal variance as a function of $q$ using Eq. \ref{eq:2.4} as a relation between the non-extensivity and fractality with a vertical line at $q=5/3$.}
\label{fig7c}
\end{figure}
%%%%%%%%%%%%%bild%%%%%%%%%%%%%%%%%%%%

%%%%%%%%%%%%%bild%%%%%%%%%%%%%%%%%%%%
\begin{figure}[tbp]
\begin{center}
\includegraphics[width=10cm, height=8cm]{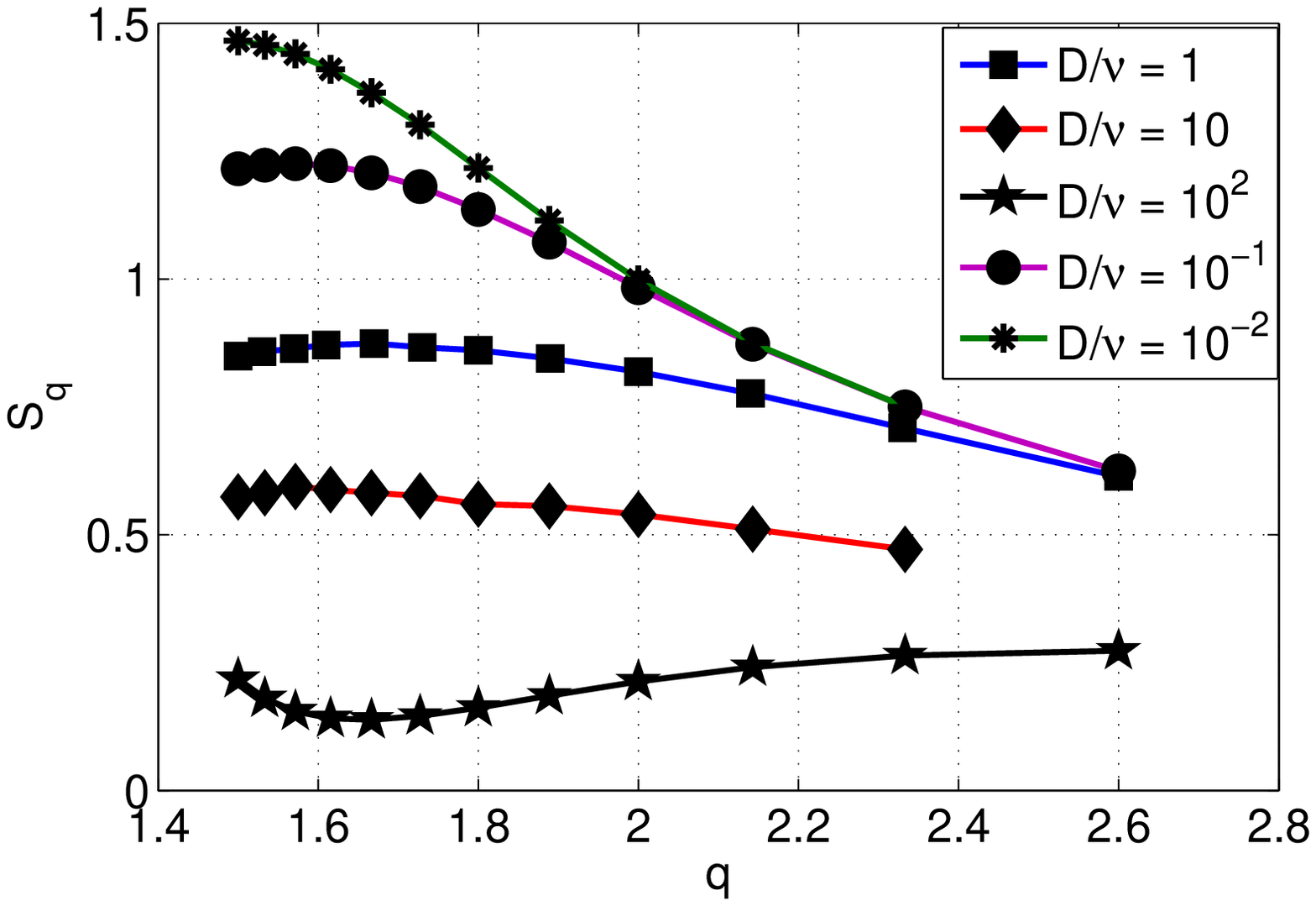}
\end{center}
\caption{The q-entropy as a function of $q$ using Eq. \ref{eq:2.4} as a relation between the non-extensivity.}
\label{fig8a}
\end{figure}
%%%%%%%%%%%%%bild%%%%%%%%%%%%%%%%%%%%

%%%%%%%%%%%%%bild%%%%%%%%%%%%%%%%%%%%
\begin{figure}[tbp]
\begin{center}
\includegraphics[width=10cm, height=8cm]{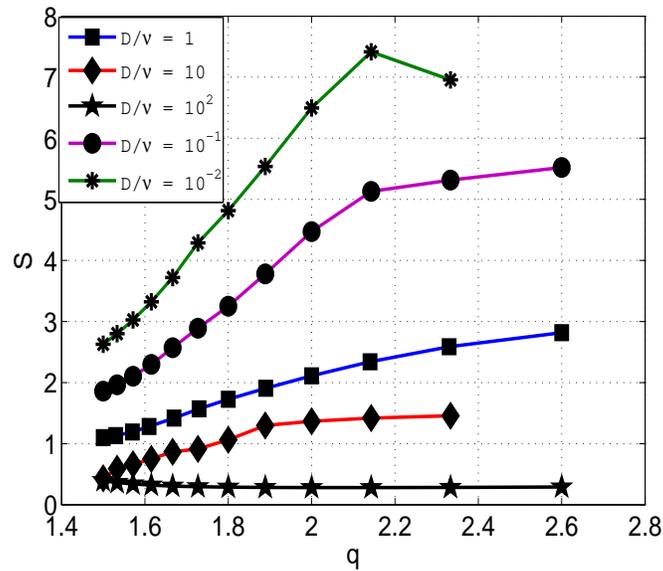}
\end{center}
\caption{The Boltzmann-Gibbs entropy as a function of $q$ using Eq. \ref{eq:2.4} as a relation between the non-extensivity and fractality.}
\label{fig8b}
\end{figure}
%%%%%%%%%%%%%bild%%%%%%%%%%%%%%%%%%%%

%%%%%%%%%%%%%bild%%%%%%%%%%%%%%%%%%%%
\begin{figure}[tbp]
\begin{center}
\includegraphics[width=10cm, height=8cm]{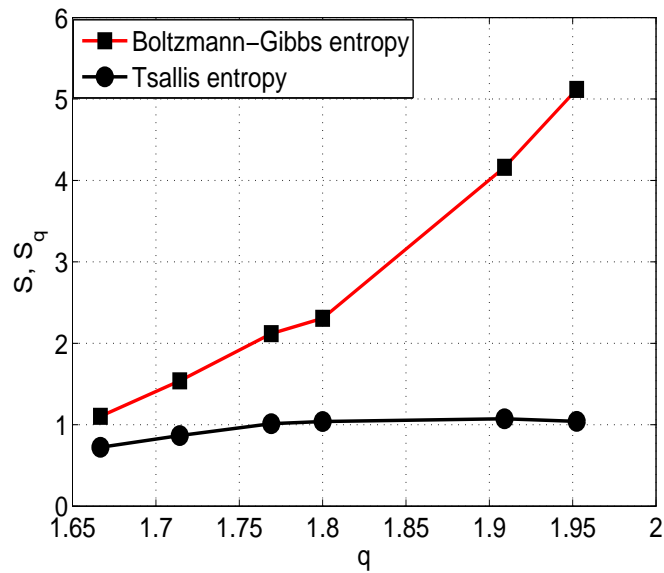}
\end{center}
\caption{The numerically found Boltzmann-Gibbs and Tsallis entropies as a function of $q$ using Eq. \ref{eq:2.4} as a relation between the non-extensivity and fractality.}
\label{fig9}
\end{figure}
%%%%%%%%%%%%%bild%%%%%%%%%%%%%%%%%%%%

\end{document}